\documentclass[superscriptaddress]{revtex4-2}
\usepackage{graphicx} % Required for inserting images

\usepackage[utf8]{inputenc}
\usepackage[cyr]{aeguill}
\usepackage{xspace}
\usepackage{amssymb}
\usepackage{physics}
\usepackage{siunitx}
\usepackage{nameref}
\usepackage{lmodern}
\usepackage{pifont}
\usepackage{subcaption}
\usepackage{comment}
\usepackage[font=footnotesize]{caption}

\begin{document}

\title{Collective adsorption of pheromones at the water-air interface}

\affiliation{ICSM, CEA, CNRS, ENSCM, Univ. Montpellier, Marcoule, France}

\author{Ludovic Jami}
\email{jami.ludovic@gmail.com}
\affiliation{Aix Marseille Univ, CNRS, Centrale Med, IRPHE (UMR 7342), Marseille, France}
\author{Bertrand Siboulet}
\affiliation{ICSM, CEA, CNRS, ENSCM, Univ. Montpellier, Marcoule, France}
\author{Thomas Zemb}
\affiliation{ICSM, CEA, CNRS, ENSCM, Univ. Montpellier, Marcoule, France}
\author{Jérôme Casas}
\affiliation{Institut de Recherche sur la Biologie de l’Insecte, UMR 7261, CNRS-Université de Tours, Tours, France}
\author{Jean-François Dufrêche}
\email{jean-francois.dufreche@icsm.fr}
\affiliation{ICSM, CEA, CNRS, ENSCM, Univ. Montpellier, Marcoule, France}

\begin{abstract}
    Understanding the phase behaviour of pheromones and other messaging molecules remains a significant and largely unexplored challenge, even though it plays a central role in chemical communication. Here, we present all-atom molecular dynamics simulations to investigate the behavior of bombykol, a model insect pheromone, adsorbed at the water–air interface. This system serves as a proxy for studying the amphiphilic nature of pheromones and their interactions with aerosol particles in the atmosphere. Our simulations reveal the molecular organization of the bombykol monolayer and its adsorption isotherm. A soft-sticky particle equation of state accurately describes the monolayer’s behavior. The analysis uncovers a two-dimensional liquid–gas phase transition within the monolayer.  Collective adsorption stabilises the molecules at the interface and the calculated free energy gain is approximately $2\:k_\mathrm{B}T$. This value increases under lower estimates of the condensing surface concentration, thereby enhancing pheromone adsorption onto aerosols. Overall, our findings hold broad relevance for molecular interface science, atmospheric chemistry, and organismal chemical communication, particularly in highlighting the critical role of phase transition phenomena.
\end{abstract}

\maketitle

\section{Introduction}

Animals—particularly insects—as well as plants, communicate through the exchange of chemical mediators. Despite the ubiquity of molecular phase exchanges in chemical communication, the physico-chemical behavior of these mediating molecules, commonly referred to as semiochemicals, remains poorly understood. Beyond the documented statistical correlations between molecular weight, volatility, semiochemical function (e.g., sexual availability, species recognition), and habitat characteristics (e.g., temperature, humidity, oxidative stability) \cite{alberts_constraints_1992}, the specific mechanisms of phase exchange involved in chemical communication are still largely unknown. Compared to other biologically or industrially relevant molecules, semiochemicals and pheromones have received relatively little attention in physical chemistry studies. Yet, a deeper comprehension of their phase change behavior is crucial for elucidating their ecological and signaling roles.

Moth pheromones are the first discovered examples and model case studies of chemical communication. Successful mate attraction in moths critically depends on the long-range persistence of these pheromonal signals. These pheromones typically consist of linear molecules with carbon chains ranging from 10 to 18 atoms and molar masses between 200 and 400 g/mol \cite{ando2004lepidopteran}. They have a low volatility, comparable to the volatility  of “base” notes in the formulation of fragrance mixtures. The bombykol, main sexual pheromone ((E, Z)-10,12-hexadecadienol) is a single chain fatty alcohol and has a saturation vapor pressure of the order of 1 mPa \cite{Fieber}. The vapor pressure of fatty alcohols from pentanol has been studied experimentally versus temperature \cite{Nasirzadeh}. Extrapolation of the vapor pressure to 18 carbon atoms without taking into account the double bound, leads to a vapor pressure of bombykol of the order of 1 mPa indeed. This is again consistent with the value obtained using the appropriated software COSMO-RS \cite{Fieber}. Many moth pheromones exhibit polar functional groups, often incorporating one or two oxygen atoms. As a result, these molecules are amphiphilic, possessing both hydrophilic and hydrophobic regions. This structural amphiphilicity suggests they may exhibit surface activity and could participate in the formation of complex structures such as micelles or emulsions. It is therefore essential to investigate their amphiphilic behavior to better understand the mechanisms involved at the various stages of chemical communication—from storage in the secretory glands, release and transport through the air, deposition on the antennae, and migration through the lymph fluid of olfactory pores, to ligand-receptor interaction with odorant-binding proteins. In particular, during atmospheric transport, pheromones—especially in the case of moth pheromones, which combine low volatility and amphiphilic properties— may not be simply transported in the gas phase, as generally assumed, but may interact with aerosols, airborne particles naturally present in the atmosphere. Understanding gas–particle phase exchanges is furthermore relevant given that some insects are known to emit pheromones directly in aerosol form \cite{krasnoff_sex_1988, yin_sex_1991, kuba_production_1988}. 

We recently proposed a mesoscopic model \cite{Jami2020} showing that most pheromones can leave the gas phase relatively quickly during chemical communication processes if the individual adsorption free energy of the molecules is high. The molecules then settle on the surface of aerosols and are delivered to the receptor in clusters. In the case of bombykol on ideal aqueous aerosols, molecular dynamics simulations \cite{Jami2022} have shown that the free adsorption energy of a single molecule is not sufficient to allow individual adsorption of molecules. However, since this is a collective phenomenon, the possibility of collective adsorption must also be taken into account, as interactions between pheromones on the surface of aerosols can stabilize adsorption. Numerically, the addition of lateral interactions with other bombykol molecules may indeed lead to a gain of more than 1.0~$k_\mathrm{B}T$.molecule$^{-1}$, as can be understood by the following argument. The surface tension of pure bombykol has been studied recently \cite{Tiryaeva_2024}. Experiment and theory converge to a value of  30 mJ.m$^{-2}$ for bulk bombykol. Knowing the area per chain of 0.18 nm$^2$.molecule$^{-1}$ of a dense layer, the cohesion energy of bulk material can be estimated at first approximation as the product of surface tension by the distance between adjacent chains as 6~10$^7$~J.m$^3$ or 7 $k_\mathrm{B}T$.molecule$^{-1}$ \cite{Israelachvili}. We get a much higher estimation when using the enthalpy of vaporization, which include the effect of oriented hydrogen bonds between alcohol groups. The enthalpy of vaporization can be deduced at first approximation from the vapor ($C_v$) and liquid ($C_l$) concentrations, via the expression $RT \ln (C_v/C_l)$, where $C_v$ is 2.4~10${^{17}}$~molecules.m$^3$ and $C_l$ is 2.16~10${^{27}}$~molecules.m$^3$. This gives a vaporization enthalpy in the bulk of pure bombykol of 60 kJ.mol$^{-1}$. If the bombykol molecule is assumed to have 4 first neighbors in the bulk and only 3 at the surface of pure bombykol, the enthalpy binding a single molecule surface layer of bombykol is of the order of 45 kJ.mol$^{-1}$, \textit{i.e.} 18~$k_\mathrm{B}T$.molecule$^{-1}$. Adding 10~$k_\mathrm{B}T$ corresponding to hydrogen bonds of the alcohol group with water, we indeed obtain a binding energy between 25 and 30~$k_\mathrm{B}T$.molecule$^{-1}$ for the surface film in equilibrium with bulk gas phase.

Thus, it is quite possible that collective adsorption is favoured over individual adsorption and that, ultimately, most molecules are located on the surface of aerosol droplets. The aim of this paper is to study this hypothesis using molecular simulations. Here, we present all-atom molecular dynamics simulations of a monolayer of bombykol at the air–water interface to investigate its surface activity. Bombykol is a well-known model pheromone of Lepidoptera, and our simulations aim to characterize the phase behavior of its monolayer. We establish its Langmuir adsorption isotherm and clarify how collective interactions within the monolayer influence the vapor–adsorbed chemical equilibrium. What interests us here is actually the lateral equation of state \cite{LEONTIDIS20071591, Lunkenheimer1992, BAUDUIN20149, Ruckenstein1998} of the molecule at the water/air interface.  This is related to the adsorption isotherm \cite{IVANOV2010118} because it basically represents the interactions between surfactant molecules at the interface. Thus, the fundamental description of adsorption phenomena developed here also holds broader relevance in the field of interfacial science. Our work is also of relevance in the field of atmospheric science as exchange molecule between gas and aerosols phase can be governed by adsorption isotherms \cite{donaldsonInfluenceOrganicFilms2006, ruehl_interfacial_2016}. 

Various molecular simulation approaches have already been used to quantify sorption phenomena. Molecular dynamics (MD) approaches to adsorption, especially those using coarse-grained force fields \cite{olzynska_mixed_2016, panzuela_molecular_2019}, are widely employed. However, all-atom simulations of adsorption processes are relatively recent and remain sparse. Most existing studies focus on condensed monolayers of phospholipids—important in lung alveoli—ionic surfactants such as sodium dodecyl sulfate, or non-ionic surfactants like polyoxyethylenes used in foams and detergents \cite{abranko2013immersion, chanda2005molecular, duncan_molecular_2011, baoukina_pressurearea_2007, olzynska_mixed_2016, panzuela_molecular_2019}. Typically, these studies simulate only one or a few surface concentrations and comprehensive simulations of Langmuir isotherms remain rare \cite{baoukina_pressurearea_2007, olzynska_mixed_2016}. Surfactant monolayers are known to exhibit a wide variety of phase behaviors \cite{kaganer_structure_1999}, stemming from their complex molecular structures and the multiple possible degrees of molecular ordering. Common phase transitions include the two-dimensional gas–liquid transition, characterized by a sharp change in surface concentration with surface tension (analogous to condensation), and the transition from a disordered to an ordered liquid, characterized by two-dimensional crystallinity and parallel alignment of hydrophobic tails. Modeling these transitions at the molecular level remains a significant scientific challenge due to the slow dynamics and strong spatial correlations involved—particularly those driven by nucleation phenomena and geometric frustration \cite{baoukina_pressurearea_2007, panzuela_molecular_2019}. Our study thus constitutes a state-of-the-art contribution to the modeling of two-dimensional phase behavior in amphiphilic systems, with implications in semiochemical research and beyond.
\clearpage

\section{Methods}
\label{sec:met}

\subsection*{Thermodynamics of interfacial systems}

Pheromones with an amphiphilic structure adsorb at the water-air interface. In sufficiently high concentrations, they interact laterally with each other. The state of the 2D phase thus formed is characterised by the temperature $T$, the surface tension of the interface $\gamma$ and the number of adsorbed molecules $N_a$. From surface thermodynamics theory \cite{Rowlinson} one can define an interfacial free energy $F_a$, which is an excess term with respect to the bulk free energies of the two phases in contact with the surface. At constant temperature, the differential reads:
\begin{equation}
    \mathrm{d}F_a=\mu_a \dd N_a +\gamma \dd S
\end{equation}
where $\mu_a$ is the chemical potential of pheromone molecules adsorbed to the surface, $N_a$ their total quantity and $S$ the interface area. This equation assumes that the Gibbs plane separating the two phases is placed in such a way that the quantity of solvent (water) at the interface is zero. The Gibbs equation then links variations in chemical potential to variations in surface tension ($\gamma$):

\begin{equation}
N_a\dd \mu_a = -S \dd \gamma
\label{eq:gibbs}
\end{equation}

The Gibbs equation reflects the extensivity of the system along the lateral dimension. It gives a variational link between the variation of the chemical potential and the variation of the surface tension. This equation is valid regardless of the procedure chosen to carry out these variations at a constant temperature. Thus, if we vary the surface area $S$ or the number of adsorbed molecules $N_a$ for the same variation in surface concentration $\Gamma_a = \frac{N_a}{S}$, we will obtain the same effect on the state variables $\gamma$ and $\mu_a$. These state variables characterise the state of the system independently of its history.

Experimentally, the Gibbs equation (or its volume equivalent, the Gibbs-Duhem equation) is commonly used to obtain the chemical potentials of particles for which a direct measurement is not possible. We propose here to do the same for molecular simulations. Unlike chemical potential, surface tension is a quantity that can be calculated directly in molecular simulations of interfacial systems, and it can be expressed directly from the coordinates and velocities of the atoms. Thus, if one simulates interfaces with varying concentrations of bombykol molecules on the surface to calculate the surface tension, one can, by integrating the Gibbs equation, deduce the chemical potential of the adsorbed molecules and thus estimate the collective contribution to adsorption. The curve characterising the variation in surface tension as a function of the surface concentration of adsorbed species is the adsorption isotherm.

\subsection*{Simulated system and computation of the surface tension}

\begin{figure}[h]
\centering
\includegraphics[width=0.5\textwidth]{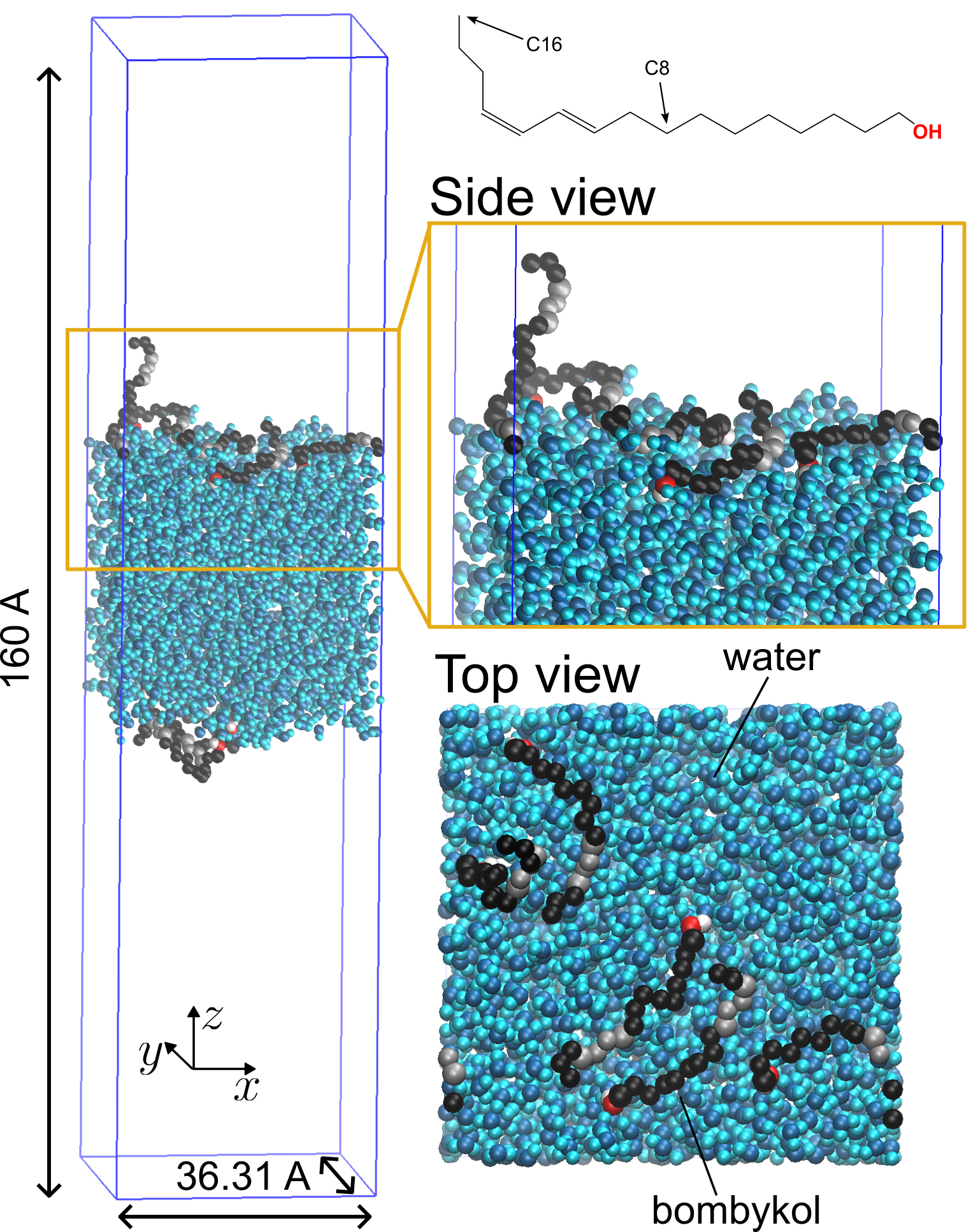}
\caption{Illustration of the simulated system of bombykol molecules at the air water interface. OH is the oxygen atom of the alcohol head group of the bombykol, C8 is the carbon atom at the center of its hydrophobic chain and C16 is its terminal carbon atom. In molecular dynamics snapshots, saturated carbons atoms are represented in black and unsaturated carbons in grey. Oxygens and hydrogens of the alcohol group are drawn in red and in white respectively. Oxygens and hydrogens of the water molecules are respectively represented in dark and light blue.} 
\label{fig:box}
\end{figure}

We simulate a slice of water made up of 1642 water molecules in a periodic box of width $L_x = 36.31\,\AA$ and height $L_z=160\,\AA$. The molecular simulation code was Lammps\cite{thompson_lammps_2022} and the simulation box is represented in Figure \ref{fig:box}. The slice of water is made up of approximately 12 layers. On each side of this slice of water, between $N_a=0$ and $N_a=40$ bombykol molecules are added, corresponding to a surface concentration varying from 0 to 3 molecules/nm$^2$.

The force fields, thermostats, and time step used are the same as for the individual pheromone adsorption simulations presented in the reference \cite{Jami2022}. It should be noted that the surface tension of systems simulated by MD is particularly sensitive to the size of the system\cite{orea_oscillatory_2005}. This must be large enough to avoid any artefacts. The simulation of Lennard-Jones (LJ) liquids suggested the following recommendations to avoid size effects: (1) $L_x > 11\sigma$ with $L_x$ the width of the interface and $\sigma$ the distance at which the LJ potential is minimal; (2) $L_z > 60\sigma$ where $L_z$ is the height of the box; (3) $L_\mathrm{slab} > 10\sigma$ where $L_\mathrm{slab}$ is the width of the liquid slice. These size effects are partly due to the truncation of LJ potentials between atoms. Adding truncation corrections to the $\gamma$ calculation for homogeneous systems (pure water, pure organic liquid), for which the density profile can be fitted by a cotangent function, or taking into account these long-range interactions in the equations of motion by lattice summation methods, may allow a correct estimate of $\gamma$ for smaller systems. Nevertheless, a simple truncation correction cannot be applied to our systems because the presence of adsorbed molecules results in density profiles that cannot be simply adjusted. Explicitly taking long-range LJ interactions into account in the simulations gives results consistent with \textit{a posteriori} corrections in the case of water\cite{vega_surface_2007, alejandre_surface_2010} but the introduction of long-range interactions artificially increases the cohesion of liquids for which no correction to the truncation of the LJ potentials has been taken into account in the parametrization of their force field \cite{fischer_properties_2015}. Long-range LJ interactions change not only the surface tension but also all properties of the liquid (for example, the enthalpy of formation). Taking these into account would make our calculations inconsistent with our previous work about individual adsorption\cite{Jami2022}. In the case of the OPLS force field, which was parametrized with a truncation correction, taking into account long-range LJ interactions did not lead to significant improvements when comparing several experimental and simulated properties of different pure organic liquids\cite{fischer_properties_2015, zubillaga_surface_2013}.

No SHAKE algorithm has been used here, thus no bonds and angles between atoms are forced to be constant. The thermalization procedure follows that used previously. Initially, a small time step is used, and is then gradually increased until $1\,$fs is reached. The velocity of the particles is meanwhile regularly re-allocated randomly so that the system is at a temperature of $\SI{25}{\celsius}$.  An attraction force of all the molecules towards the center of the box was added in order to force the initial organization of the water into a slice and that of the bombykol molecules into a monolayer at the interface. This potential is gradually reduced during thermalization. We also simulated this system for 10 ps without imposing any external forces. Our different molecular systems are then simulated for 100$\,$ns and analysed.

Surface tension was calculated from the diagonal components of the stress tensor. For a bulk solution, the pressure can be calculated using the virial theorem:\cite{hansen2014theory} :
\begin{equation}
P = \frac{Nk_BT}{V} + \left\langle \sum_i \mathbf{r}_i\sum_j\mathbf{F}_{ij} \right\rangle
\end{equation}
$N$ is the number of atoms in the system. $\mathbf{r}_i$ is the position of atom $i$. $\mathbf{F}_{ij}$ is the force that particle $j$ applies to $i$. This relation has been generalised for slab interfaces. In this case, we have to distinguish the three diagonal components of the stress tensor defined by \cite{parker_tensor_1954} :
\begin{equation}
P_{\alpha \beta} = \left\langle \mathcal{P}_{\alpha \beta} \right\rangle = \left\langle \sum_i m_i v_{i}^\alpha v_{i}^\beta+ \sum_i r_{i}^\alpha \sum_j F_{ij}^\beta \right\rangle /V
\label{eq:P_IJ}
\end{equation}
where $\alpha$  and $\beta$ denote the different directions $(x,y,z)$ of the Cartesian reference frame.  The force exerted by surface tension makes the two longitudinal components $x$ and $y$ different from the normal component $z$. This gives us (see for example\cite{vega_surface_2007}):
\begin{equation}
\gamma = \frac{L_z}{2}\left\langle \mathcal{P}_{zz} - \frac{1}{2}(\mathcal{P}_{xx} + \mathcal{P}_{yy})\right\rangle
\label{eq:Ptensor2gamma}
\end{equation}
By calculating the diagonal components of the stress tensor $\mathcal{P}_{\alpha \alpha}$ of the box every 100$\,$fs and using the equation (\ref{eq:Ptensor2gamma}), we can calculate the surface tension of our systems. Note that the $1/2$ factor is due to the presence of two water-air interfaces in the simulation box. $\gamma$. Calculating $\gamma$ for different pheromone surface concentrations allows us to deduce, by thermodynamic integration, the chemical potential of the adsorbed phase.

It should be noted that the calculation of the surface tension in these systems requires a particularly long simulation time. The surface tension fluctuates greatly in MD systems and only long simulations allow to minimise its error. The error in surface tension was estimated using the block method. The surface tension is averaged over blocks of sizes between 1 and $2500$ fs, and the maximum error obtained for these different block sizes is chosen as the error on $\gamma$. In principle, this method gives a larger estimate of the error, but in practice this is very limited ($<1\%$).

\subsection*{Equations of state (EOS) to describe monolayer isotherms}

There are several ways to relate surface tension to chemical potential by integrating Gibb's equation. This can be done numerically, but the lack of data, particularly for low concentrations requiring the simulation of a large number of water molecules, means that it is more accurate to use an analytical equation of state. We considered two equations of state (EOS),  the Smith–Ivanov–Ananthapadmanabhan–Lips (SIAL) EOS and a  soft-stickyparticule EOS \cite{IVANOV2010118}. The SIAL EOS combines the nearly exact equation of Helfand–Frisch–Lebowitz (HFL) for two-dimensional hard disc fluid with an attractive mean field term $-a\Gamma_a^2$ \cite{slavchov_adsorption_2017}:
\begin{equation}
    \gamma_\mathrm{w}-\gamma = k_BT\frac{1/\Gamma_a}{\left(\frac{1}{\Gamma_a}-b\right)^2}-a\Gamma_a^2
\label{eq:SIAL}
\end{equation}
$\gamma_\mathrm{w}$ is the surface tension of pure water. The parameter $b$ is the hard disc area of the molecule quantifying the steric repulsion between the adsorbed molecules. The soft-sticky particle EOS considers the same mean field attractive term $-a\Gamma_a^2$  but a softer repulsive term taken from the soft-particles ($n = 6$) EOS \cite{tan_virial_2011} :
\begin{equation}
    \gamma_\mathrm{w}-\gamma = k_BT\Gamma_a Z(\sigma^2\Gamma_a)-a\Gamma_a^2
\label{eq:soft-sticky-particules}
\end{equation}
Where $Z(\sigma^2\Gamma_a)=Z(\rho)$ is the compressibility factor associated to the repulsion of the soft particles interacting with a two body potential, $w^{(2)}(r)=k_{\mathrm{B}}T\qty(\frac{\sigma}{r})^6$, diverging as power 6 with the distance $r$ between them and characterised by the repulsion distance $\sigma$. The Padé 5-3 approximant of this compressibility factor has been computed by Tan, Shultz and Kofke and reads \cite{tan_virial_2011}:
\begin{equation}
    Z(\rho) = \frac{\begin{matrix}1.0 + 7.432255\rho + 23.854807\rho^2 + 40.330195\rho^3 \\+ 34.393896\rho^4 + 10.723480\rho^5 \end{matrix}}{1.0 + 3.720037\rho + 4.493218\rho^2 + 1.554135\rho^3}
    \label{eq:soft-particules}
\end{equation}
    
Both models share the same physical characteristics. For sufficiently large attraction parameter $a$ or sufficiently small temperature or repulsion ($b$ or $\sigma$), the SIAL and the soft-sticky particles EOS display a 2D liquid-gas phase transition. The difference comes from the soft repulsion of soft-sticky particles model, which allows for greater compressibility at high density.

More broadly, the EOS here considered both fall in the category of particle based model like the classical Van der Waals EOS. Notably such models are commonly used to describe phase transition in monolayers of non-ionic surfactants\cite{slavchov_adsorption_2017, lewandowski_adsorption_2019}. They are used in particular to describe the hygroscopic properties of aerosols containing organic molecules with low solubility and/or surface activity. Depending on the humidity and therefore on the water that condenses on the aerosol, these organic molecules can pass from a gaseous phase to a adsorbed phase as a dilute or dense monolayer with important consequences on the quantity that can condense\cite{ruehl_surface_2014, ruehl_interfacial_2016}.

\subsection*{Maxwell construction for the 2D phase separation}

The possibility of having a two-dimensional liquid/gas phase separation makes it necessary to modify the equation of state in the coexistence domain by a Maxwell construction. The procedure used here is as follows. If the 2D phase separation occurs, it leads to a characteristic plateauing of the $\gamma(\Gamma_a)$ isotherm curve. Let us consider the surface concentrations for which the 2D gas and liquid phases coexist. As they are in mechanical and chemical equilibrium, they share the same surface tension and the same chemical potential.  Let us note $\Gamma_{C}$ the condensing surface concentration, \textit{i.e.} the vapor concentration when the 2D liquid phase begins to appear and $\Gamma_{V}$ the vaporisation surface concentration,  \textit{i.e.} the liquid concentration when the 2D vapor phase begins to appear $\left( \Gamma_{C}< \Gamma_{V} \right) $. Let us note the corresponding chemical potentials of the two phases $\mu_{C}$ and $\mu_{V}$. Thus, under phase transition conditions, the EOS (\ref{eq:SIAL}) or (\ref{eq:soft-sticky-particules}) can no longer describe the surface tension in our system and we have for $\Gamma_a \in [\Gamma_{C}; \Gamma_{V}]$, $\gamma$ that is equal to the surface tension of saturation $\gamma_\mathrm{sat}$.\\
Lets now relates $\Gamma_{C}$, $\Gamma_{V}$ and $\gamma_\mathrm{sat}$ to the parameters of the EOS. Since the gas and liquid states can both be described by the same equation of state, the integration of the Gibbs equation \ref{eq:gibbs} links $\mu_{V}$ and $\mu_{C}$ by:
\begin{equation}
\mu_{V} = \mu_{C} - \int_{\Gamma_{C}}^{\Gamma_{V}}\frac{\dd\gamma}{\Gamma_a}
\end{equation}
At equilibrium $\mu_{V}=\mu_{C}$ so that
\begin{equation}
    \int_{\Gamma_{C}}^{\Gamma_{V}}\frac{\dd\gamma}{\Gamma_a}=0
\end{equation}
This equation corresponds to the Maxwell construction for the 2D liquid/gas equilibrium. The integration by part of this condition imposes for the surface tension:
\begin{equation}
\qty(\gamma_\mathrm{w} - \gamma_{\mathrm{sat}})\left(\frac{1}{\Gamma_{C}}-\frac{1}{\Gamma_{V}}\right) = -\int_{\Gamma_{C}}^{\Gamma_{V}}\qty(\gamma_\mathrm{w} - \gamma)\dd{\frac{1}{\Gamma_a}}
\label{eq:equal_area_rule}
\end{equation}
Here we used the fact that the condensing and vaporisation surface tension are equal and we note their common value $\gamma_\mathrm{sat}$.

In the case of the SIAL EOS, the condition (\ref{eq:equal_area_rule}) can be expressed explicitly as:
\begin{equation}
\begin{split}
\qty(\gamma_\mathrm{w} - \gamma_\mathrm{sat})\left(\frac{1}{\Gamma_{C}}-\frac{1}{\Gamma_{V}}\right)
={} &   k_B T \ln{\frac{1/\Gamma_\mathrm{C} - b}{1/\Gamma_\mathrm{V} - b}} -  k_B T b  \left( \frac{1}{1/\Gamma_\mathrm{C} - b} - \frac{1}{1/\Gamma_\mathrm{V} - b}\right)\\
& + a \left( \Gamma_\mathrm{C} - \Gamma_\mathrm{V} \right)
\end{split}
\label{eq:equal_area_rule_SIAL}
\end{equation}
Then, the practical calculation of the equation of state, taking account of this phase separation, was carried out as follows. First, the equilibrium quantities $\gamma_\mathrm{sat}$, $\Gamma_{C}$ and $\Gamma_{V}$ have been calculated if the phase transition occur. These values are functions for the values of the parameters $a$ and $b$ or $\sigma$ considered. In order to do this, we solved the system of equations given by the equation \ref{eq:equal_area_rule} and the EOS (\ref{eq:SIAL} or \ref{eq:soft-sticky-particules}) at the condensing and vaporisation points (i.e. by replacing $\Gamma_{a}$ by $\Gamma_{C}$ or $\Gamma_{V}$ and $\gamma$ by $\gamma_\mathrm{sat}$).  Next, $\gamma$ is calculated using the equation of state, but if $\Gamma_a \in [\Gamma_{C}; \Gamma_{V}]$, the obtained value is replaced by $\gamma_\mathrm{sat}$.

\subsection*{Activity coefficient of the adsorbed phase}

The chemical potential of the adsorbed molecules is :
\begin{equation}
\mu_a = \mu_a^o + k_\mathrm{B}T\ln\left(\gamma_a\frac{\Gamma_a}{\Gamma^o}\right)
\label{eq:mu_a}
\end{equation}
where $\Gamma_a$ is the surface concentration of adsorbed molecules. $\gamma_a$ is the activity coefficient of the adsorbed molecules (not to be confused with the surface tension $\gamma$). $\Gamma_a^o$  and $\mu_a^o$ are respectively the standard surface concentration and the standard chemical potential. The precise definition of these quantities is somewhat arbitrary \cite{Dufreche24}. It depends on the choice of the standard two-dimensional state considered. The only important thing is that, within the limits of dilution, when the molecules cannot be seen ($\Gamma_a \to 0$), Henry's law on the surface is found: $\mu_a \to A + k_\mathrm{B} T \ln \Gamma_a$.  We have considered here an infinitely dilute reference where $\gamma_a \to 1$ when $\Gamma_a \to 0$. So we have $A=\mu_a^o+k_\mathrm{B}T\ln\frac{1}{\Gamma^o}$. The standard concentration chosen for plotting the curves was arbitrarily $\Gamma^o=1/L_x^2$. Note that this choice does not change the value of $\gamma_a$.

For a sufficiently dilute monolayer, the bombykol molecules do not interact with each other:  we have $\gamma_a = 1$ and $\mu_a=\mu_a^\mathrm{ideal}=\mu_a^o + k_\mathrm{B}T\ln(\frac{\Gamma_a}{\Gamma^o})$. We then find the chemical potential of a 2D perfect gas and, using the Gibbs equation \ref{eq:gibbs}, we obtain the equation of state of an ideal monolayer:
\begin{equation}
\gamma_\mathrm{w} - \gamma = k_\mathrm{B}T\Gamma_a
\label{eq:monolayer_ideal}
\end{equation}
with $\gamma_\mathrm{w}$ the surface tension of pure water. Note here that $\gamma_\mathrm{w} - \gamma$ acts as a 2D pressure for the 2D gas of adsorbed molecules. This equation is nothing but the generalization of the ideal gas law $PV=Nk_\mathrm{B}T$ for a 2D system. 

In the general case, 
\begin{equation}
\gamma_\mathrm{w} - \gamma = k_\mathrm{B}T\Gamma_a+f(\Gamma_a)
\end{equation}
with $f(\Gamma_a) = O(\Gamma_a^2) $. The Gibbs equation (\ref{eq:gibbs})  yields $\mu_a=A + k_\mathrm{B} T \ln \Gamma_a+\int_{\Gamma_a^o}^{\Gamma_a} \frac{f^\prime(\Gamma_a)}{\Gamma_a}\:\mathrm{d}\Gamma_a$ so that:
\begin{equation}
     k_\mathrm{B}T\ln \gamma_a = \int_{\Gamma_a^o}^{\Gamma_a} \frac{f^\prime(\Gamma_a)}{\Gamma_a}\:\mathrm{d}\Gamma_a
\end{equation}
It is therefore possible to calculate the activity coefficient $\gamma_a(\Gamma_a)$ from the surface tension expression $\gamma(\Gamma_a)$.

\subsection*{Thermodynamics of adsorption into a non-ideal monolayer}
Let us relate the chemical potential of the bombykol monolayer, determined from its isotherm, to the adsorption equilibrium at the water-air interface. The idea is to make the connection with the adsorption of a single molecule on the surface, studied in reference \cite{Jami2022} to determine the free energy of adsorption of a molecule as a function of the surface concentration. Thus, we consider the phase equilibrium between the free molecules in the air and those adsorbed on the surface. It is therefore a chemical equilibrium between a 3D gas of molecules in the air with chemical potential $\mu_g$ and a 2D layer of adsorbed molecules with chemical potential $\mu_a$. The equality of the chemical potentials of bombykol in these two states then reads
\begin{equation}
\mu_a = \mu_g
\label{eq:eq_chim_a-g}
\end{equation}
The chemical potential of pheromones in the gaseous state can be described by a perfect gas model:
\begin{equation}
\mu_g = \mu_g^o + k_\mathrm{B}T\ln \left(\frac{C_g}{C^o} \right)
\label{eq:mu_g}
\end{equation}
$C_g$ being the concentration in the gas, $\mu_g^o$ the chemical potential in the standard state (which depends only on temperature), $C^o$ the concentration in the standard state (corresponding to a partial pressure $P^o=C^ok_\mathrm{B}T=1\:$bar) and $k_\mathrm{B}$ Boltzmann's constant.

The chemical potential of the adsorbed molecules is given by equation (\ref{eq:mu_a}). Injecting the equations (\ref{eq:mu_g}) and (\ref{eq:mu_a}) into the chemical equilibrium (\ref{eq:eq_chim_a-g}) gives :
\begin{equation}
K_Le^{-\ln\gamma_a}=\frac{\Gamma_aC^o}{\Gamma^oC_g}
\label{eq:ads_coll}
\end{equation}
with $K_L = \exp(-\frac{\mu_a^o-\mu_g^o}{k_BT})$ the adsorption equilibrium constant already considered in reference \cite{Jami2022}. Thus, the quantity $k_BT\ln\gamma_a = \mu_a - \mu_a^\mathrm{ideal}$ can be considered as an added free energy, due to the lateral interaction of the bombykol molecules in the monolayer, favouring the adsorption of bombykol compared with an ideal monolayer (without lateral interaction). In the previous study \cite{Jami2022}, only the ideal term was calculated. Adsorptions were thus assumed to be individual. It appeared that in the case of bombykol on aqueous aerosols, the adsorption free energy was not sufficient to allow pheromones to preferentially attach to aerosols present in the air \cite{Jami2020}. Only half of the free energy required was obtained, meaning that typically another $14\:k_\mathrm{B}T$ were needed for the pheromones to leave the air and accumulate on the surface of the aerosols.
\clearpage

\section{Results \& Discussions}

\subsection{The interaction of two bombykol molecules at the interface}

We first describe the actual interactions between bombykol molecules on the surface. To do this, we estimated the 2D mean force potential $w^{(2)}(r)$ between two molecules averaged over the water configurations, from simulations with two molecules on the surface ($r$ is a 2D distance).  Let $\mathcal{P}(r)$ denote the probability density that two molecules are at a distance $r$. and $g^{2D}_r(r)$ the 2D pair correlation function. We have the probability differential  $\mathrm{d}p=\mathcal{P}(r)\mathrm{d}r= g^{2D}_r(r)2 \pi r \mathrm{d}r=B' \exp\left( \frac{-w^{(2)}(r)}{k_\mathrm{B}T} \right)2 \pi r \mathrm{d}r$ so that
\begin{equation}
    w^{(2)}(r)=-k_{\mathrm{B}}T \ln \mathcal{P}+k_{\mathrm{B}}T \ln r + B
\end{equation}
We used this equation to evaluate $w^{(2)}(r)$, the constant $B = \ln(2\pi B')$ being chosen so that $w^{(2)}(r)\to 0$ if $r\to 0$.

\begin{figure}[h]
\centering
\includegraphics[width=0.5\textwidth]{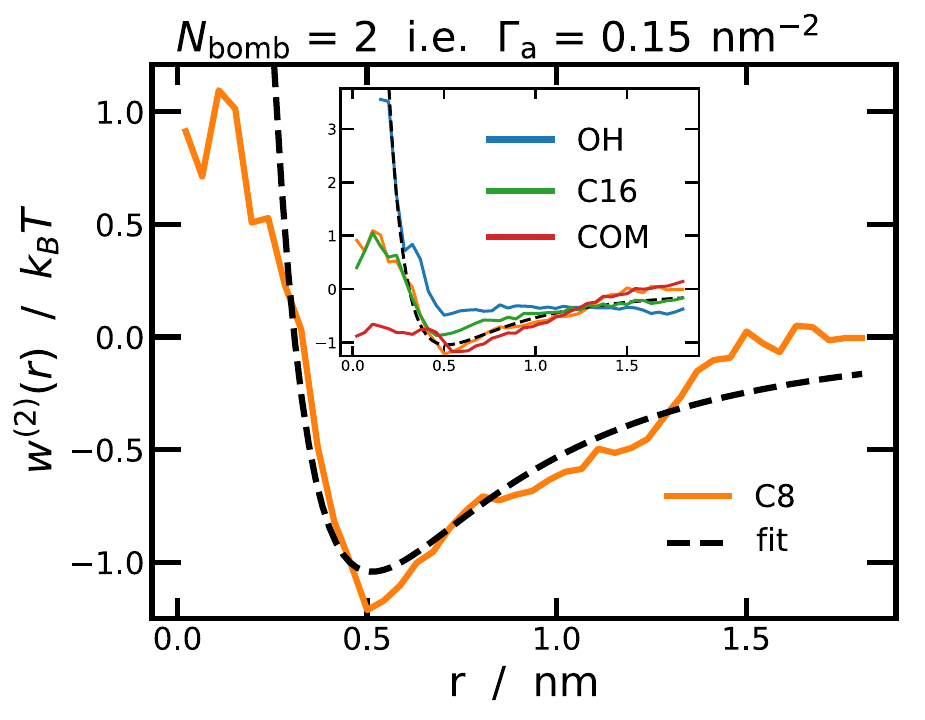}
\caption{Potential of interaction of two bombykol molecules at the water-air interface calculated for different sites of the molecule represented in figure 1: medium carbon (C8), oxygen of the hydroxyl head group (OH), terminal carbon (C16) and center of mass (COM). The dashed line represents the fitted function equation \ref{eq:fitw2}. It is calculated for different points of the bombykol molecules represented in Fig. \ref{fig:box}. COM is the center of mass.}
\label{fig:2bomb}
\end{figure}

The results are presented in figure \ref{fig:2bomb}. Several sites in molecules have been considered to define $w^{(2)}(r)$. Globally, by considering the central carbon C8, the potential is attractive over long distances ($r > 0.5\:$nm), and repulsive otherwise ($r<0.5\:$nm). The depth of the attraction is close to $k_{\mathrm{B}}T = 2.5\:$kJ.mol$^{-1}$. This suggests the possibility of liquid/gas phase separation on the surface. The curve can be fitted by the soft sphere 2D potential
\begin{equation}
        w^{(2)}(r) = \epsilon \left(\frac{\sigma}{r+r_0}\right)^6 -\epsilon \left(\frac{\sigma}{r+r_0}\right)^3 
\label{eq:fitw2}
    \end{equation}
with $\epsilon = 1.0\, k_BT$, $\sigma = 0.99\,$nm \& $r_0 = 0.48\,$nm.  The repulsion is therefore not very hard, since the power is not very high (6). Changes in the configuration of the molecules allow them to move closer together by deforming slightly. The attraction is also long-range, due to specific interactions between the chains.

We can also look at the effective potentials at other sites of the molecule (see inlet  of fig. \ref{fig:2bomb}). For OH, the attraction disappears. This is due to the position of the heads, which interact preferentially with water and therefore have little interaction with neighboring molecules. The attraction between the centers of mass is logically the same as that between the central carbons, but the repulsion disappears almost entirely. This is simply because the center of mass does not necessarily correspond to a material point on the molecule. There is no direct steric repulsion between the sites, and it is possible for two molecules to move closer together in terms of their centers of mass if the molecule deforms.

\subsection{An equation of state to capture the phase transition of bombykol at the interface}

\begin{figure}
\centering
\includegraphics[width=\textwidth]{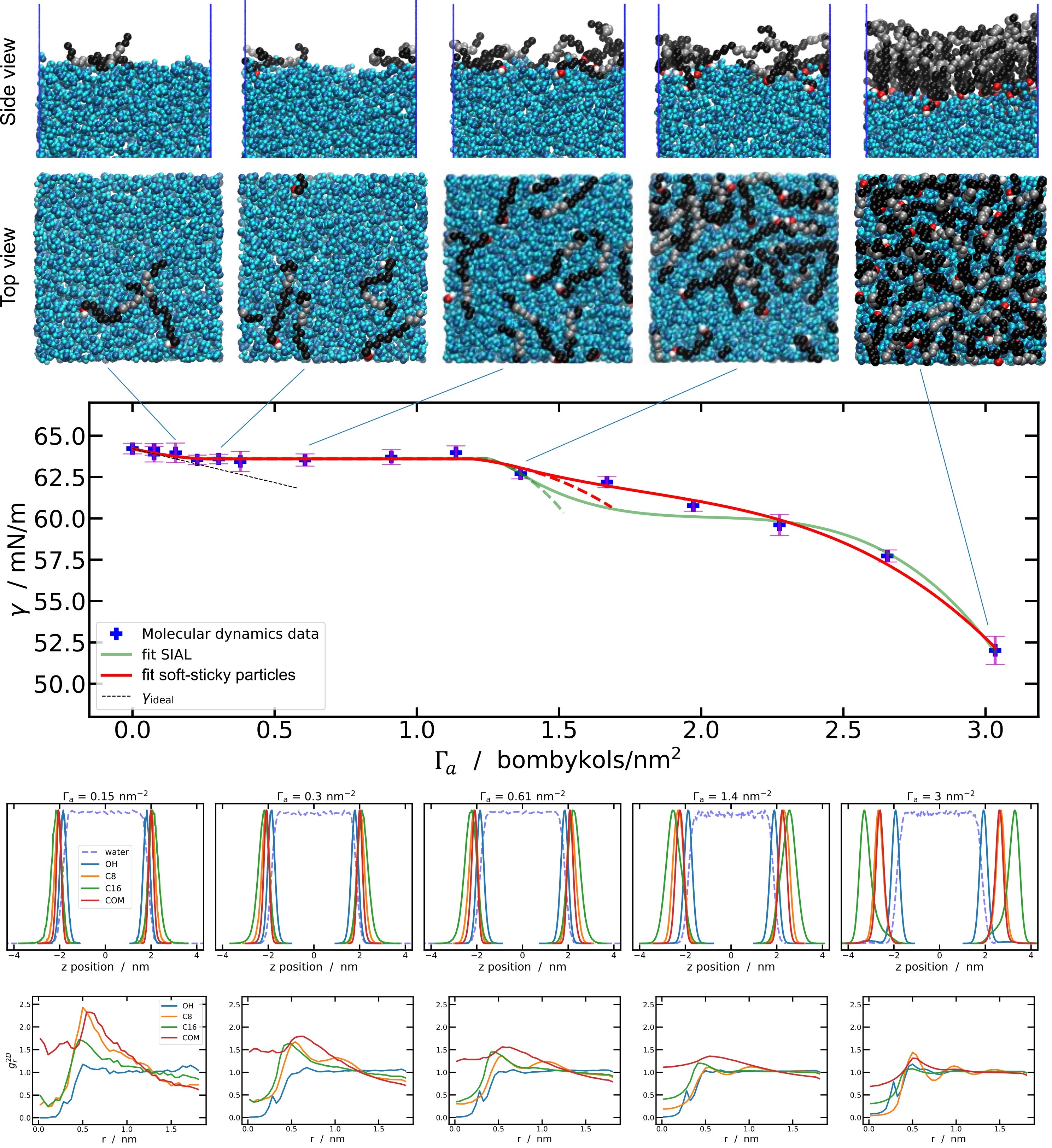}
\caption{Snapshot, surface tension (isotherm), histogram in $z$ and pair correlation functions $g_r$ of several sites in bombykols molecules. For the surface tension plot, the solid curves correspond to the overall fits, while the dashed curves correspond to the two EOS, extended for $\Gamma_a>1.4\:$nm$^{-2}$.}
\label{fig:isotherm_&_g_r}
\end{figure}

Figure \ref{fig:isotherm_&_g_r} shows the surface tension $\gamma$ as a function of $\Gamma_a$. For a few representative values of $\Gamma_a$, snapshots, $z$ histograms and pair-correlation function $g_r^\mathrm{2D}$ are given. For snapshots, we note that the molecules remain on the surface of the water and do not desorb during simulations, indicating significant free adsorption energy. At low concentrations, the carbon chains are parallel to the surface and the molecules tend to align themselves parallel to optimize attraction. Thus, even at low concentrations, interactions between molecules are significant, and they tend to move closer together. Quite quickly, significant aggregation can be seen, possibly indicating a liquid phase, but given the small size of the system, it is not easy to be sure. At high concentrations, the entire surface is saturated with molecules. A change in configuration can be observed in this collective absorption. While the molecules are parallel to the surface at low concentrations, in order to optimize interactions with water, when the surface is saturated, the molecules turn upright. Ultimately, only the polar heads interact with water. At high concentrations, the carbon chains become perpendicular to the surface, with the tails containing double bonds facing the air.

\begin{figure}
\centering
\includegraphics[width=0.5\textwidth]{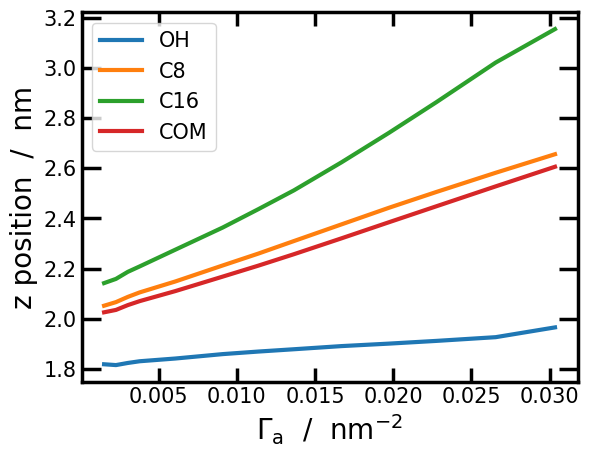}
\caption{Average z position of the bombykol atoms as a function of the surface concentration. The bombykol molecules stands more and more vertical as the surface is densily covered.}
\label{fig:posZ}
\end{figure}

This is confirmed from the histogram analysis given in the same figure. Only the polar heads remain constantly close to the water when the quantity of adsorbed molecules varies. In contrast, the other sites of the molecule gradually move away from the water, reflecting this reorientation of the molecules. For low concentration, the carbon at the end of the C16 molecule is at the same altitude as the central carbon C8 in close proximity to OH, but gradually moves away from it. It typically passes from 2 \AA  \ to almost 2~nm from the water surface. The interface also widens, reflecting an increase in surface fluctuations with decreasing surface tension. This effect is explicitly plotted in Fig. \ref{fig:posZ}. The curves are almost straight lines. This transition is therefore not sudden. There is no visible separation between a stable configuration at low concentration with parallel molecules and another stable configuration at high concentration with perpendicular molecules. Rather, it is a gradual reorientation of the molecules. However, caution is warranted here. If the molecules undergo a 2D phase transition, considering the size of the system, the molecules may switch between phases, thus fluctuations are important and it may mask a change in behavior. As soon as the concentration is not zero, the molecules begin to straighten up.

The pair correlation functions  $g^{2D}_r(r)$ (fig. \ref{fig:isotherm_&_g_r}) also depend on the surface concentration. For very low $\Gamma_a$, we recover the effective potential represented in figure \ref{fig:2bomb}, $g_r^{2D}=B'\exp\left( \frac{-w^{(2)}(r)}{k_\mathrm{B}T}\right)$.  In the case of C8, which is characteristic of molecular interactions, a second peak appears at $\Gamma_a = 0.3\:$nm$^{-2}$ indicative of  the formation of clusters. This effect becomes increasingly significant with concentration, and eventually a third peak also appears. The other sites of the molecule appear less structured due to its rotations and deformations. Globally, as the surface concentration $\Gamma_a$ increases, the intensity of the first peak decreases, with the C8 peak falling from 2.4 to 1.5, for example. It therefore appears that the interface enhances the interactions among molecules at high concentrations by aligning them favorably and thus promoting condensed phases.
 
The surface tension is also shown in the figure \ref{fig:isotherm_&_g_r} as a function of $\Gamma_a$. Overall, it decreases, but this decrease is not uniform. For very low concentrations, the results are consistent with the ideal law but the variations are within the error bars. A noticeable plateau then appears in an intermediate range, typically between 0.2 and 1.2 nm$^{-2}$. For high concentration values, on the contrary, the surface tension decreases more sharply. 

The numerical results of MD can be interpreted thanks to the SIAL and soft-sticky particle EOS (sec. \ref{sec:met}). Thus, the curve was ajusted using the two models. The procedure was as follows. The two equations of state were only used for surface concentrations below 1.4~nm$^{-2}$ because they are unable to describe the isotherm at high concentrations. In fact, these equations of state ultimately only improve and generalize the Van der Waals model for surfaces. Thus, the penetration of chains, which is modified by the reorientation phenomenon, is not taken into account. For low surface concentrations, a plateau predominates and their ability to describe phase separations means that these EOS can be used. However, at high concentrations, this chain penetration increases significantly and even becomes more important than the soft-sticky particle model predicts, as can be seen from the effective interaction potential. These EOS are therefore no longer able to describe the isotherm. Thus, for $\Gamma_a>1.4$~nm$^{-2}$, the EOS were replaced by a polynomial:
\begin{equation}
    \gamma_\mathrm{w} - \gamma = C_0+C_1 \Gamma_a + C_2 \Gamma_a^2 +C_3 \Gamma_a^3
\end{equation}
The values of the $C_i$'s were adjusted from the molecular dynamics points, taking into account the condition that the curves should have a continuous derivative at $\Gamma_a=1.4$~nm$^{-2}$. The values then obtained from the fit of the MD data are: $C_2 = 0.5772\:$nm$^{4}$.mN/m \& $C_3 = -201.4\:$nm$^{6}$.mN/m for the SIAL model and $C_2 = 0.2371\:$nm$^{4}$.mN/m \& $C_3 = -52.98\:$nm$^{6}$.mN/m for the soft-sticky particle model.  Globally, we observe in the figure \ref{fig:isotherm_&_g_r} that this extension is necessary because in simulations, surface tension decreases less rapidly than predicted by the equations of state. In fact, two-dimensional repulsion becomes increasingly soft at high concentrations when molecules become upright. This therefore reduces the effective steric repulsion between molecules. The molecules therefore attract each other more, and the surface tension, which reflects the system's tendency to reduce its surface area, remains higher than predicted by the EOS. The soft-sticky particle EOS which models a soft repulsion performs better than the SIAL EOS which is based on a hard disk fluid model.

The following parameters for the EOS are obtained. For the SIAL EOS, $a=9.12\ 10^3 \ $pN.\AA$^{3}$ and $b=0.3289\ $nm$^{2}$. For the soft-sticky particle EOS,  $a=1.30\ 10^4 \ $pN.\AA$^{3}$ and $\sigma=0.615\ $nm. These results are globally quite consistent since both the attraction terms $a$ and the size terms $\sqrt{b}$ and $\sigma$ are close. The size terms are still about two times bigger than the distance that minimize the Lennard-Jones potential ($\sigma_\mathrm{LJ-OH}/0.89 = 3.5$) describing the interaction of the oxygen of the alcohol head group with the other atoms. Indeed at high concentration, the alcohol head group tends to be separated by one water molecule. The areas obtain from isotherm fitting is also about two time larger than the hard disk area $\alpha = 0.165\,$nm$^2$ obtained from crystallographic measures of alcohol series\cite{slavchov_adsorption_2017}. This is also indicating that bombykol alcohol head groups are not found closed packed on the water surface. The value of the attraction terms $a$ can be compared to the one obtained from the effective potential $w^{(2)}(r)$. From the attractive term of the fitted 2D potential (\ref{eq:fitw2}), we can calculate the mean field value of $a =  - \frac{1}{2}\int w^{(2)}_\mathrm{attr}(r)2\pi r\mathrm{d}  r = \frac{3\pi\epsilon\sigma^3}{4r_0} =1.96\ 10^4 \ $pN.\AA$^{3}$. Here again, the results are consistent, even if the mean field approximation overestimate the actual cohesion of concentrated molecules. 

The associated surface concentration values are as follows.  For the soft-sticky particle EOS $1/\Gamma_\mathrm{V} = 83.166$ \AA$^2$ and  $1/\Gamma_\mathrm{C} = 408.206$ \AA$^2$ and for the SIAL EOS $1/\Gamma_\mathrm{V} = 80.043$ \AA$^2$, $1/\Gamma_\mathrm{C} = 492.813$ \AA$^2$. The liquid phase therefore appears rapidly at the surface, \textit{a priori} when the average distance between the molecules is greater than $1/\sqrt{\Gamma_\mathrm{V}}\approx 2\ $nm.

\begin{figure}
\centering
\includegraphics[width=0.50\textwidth]{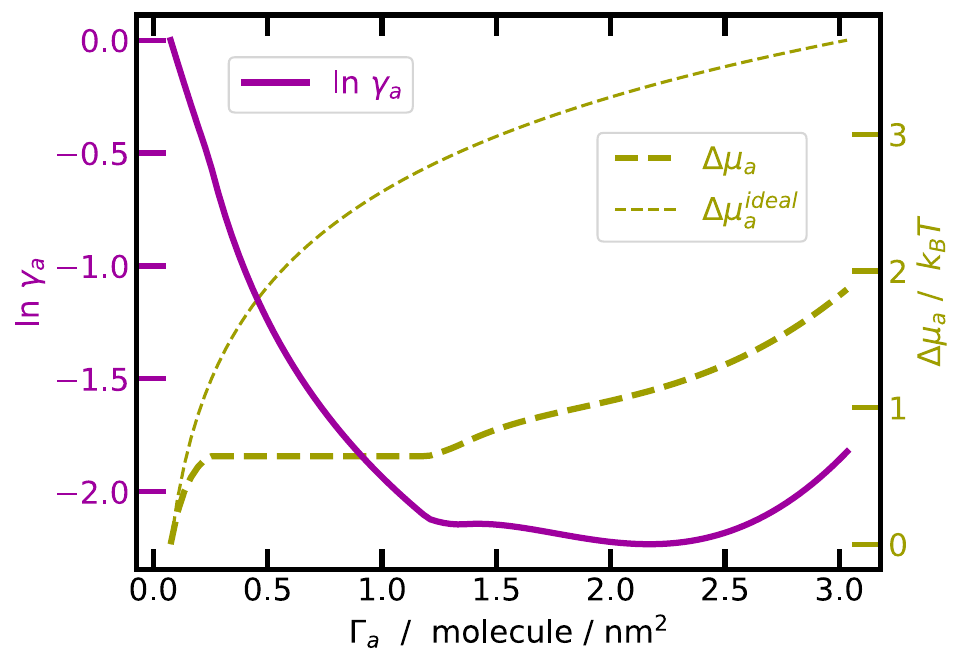}
\caption{Chemical potential (left) and activity coefficient (right) in $k_\mathrm{B}T$ unit of a monolayer of bombykol as a function of the surface concentration $\Gamma_a$. Since the chemical potential is only known to within an additive constant, the standard term has been removed: $\Delta \mu_a = \mu_a - \mu_a(\Gamma_a=\frac{1}{L_x^2})$ and $\Delta \mu_a^{\mathrm{ideal}} = k_\mathrm{B}T \ln \left( \Gamma_a L_x^2 \right)$ }
\label{fig:activite_bomb_ads_soft_sphere}
\end{figure}

The chemical potential and the associated activity coefficient can then be calculated as described in section \ref{sec:met}. The results are plotted in figure   \ref{fig:activite_bomb_ads_soft_sphere}  using the soft-sticky particle EOS model to fit the bobmykol isotherm. The chemical potential deviates from the ideal chemical potential as soon as the liquid phase appears. It remains constant throughout the entire coexistence range between the two phases. It then increases again when all molecules are condensed. The activity coefficient decreases sharply at low surface concentrations and when phase separation occurs. However, as soon as the two phases no longer coexist, it becomes more or less constant and even rises again at very high saturations. The attraction between molecules, which is very visible in the simulations, thus greatly reduces the activity coefficient and therefore the adsorption free energy. This decrease appears at the beginning in the dilute regime where the molecules attract each other. However, once the molecules have completed their condensation, the effective forces switch to a repulsive regime and the free energy no longer decreases.

Quantitatively, the activity coefficient term $k_\mathrm{B}T \ln \gamma_a$ goes down to $2 \ k_\mathrm{B}T$. This value is therefore insufficient to permit, under realistic conditions, the adsorption of pheromones on particulate phase \cite{Jami2020, Jami2022} since, as indicated in the previous section, approximately $14\ k_\mathrm{B}T$ is required. The processing of the molecular dynamics data proposed here does not therefore support the favorable adsorption of bombykol on pure aqueous aerosols. However, we must be very cautious with this conclusions for the following important reason. In fact, the value of the activity coefficient is completely determined by $\Gamma_\mathrm{C}$, the condensing surface concentration, \textit{i.e}. the vapor concentration when the 2D liquid phase begins to appear.  This can be understood by the following reasoning. When phase separation occurs, the chemical potential is constant. It is therefore equal to that of the gas phase. If the gas phase is very dilute, the activity coefficient correction can be neglected. The chemical potential is then constant for $\Gamma_a \in [\Gamma_{C}; \Gamma_{V}]$ and its value reads:
\begin{equation}
    \mu_a=\mu_a^o+k_\mathrm{B}T \ln \frac{\Gamma_\mathrm{C}}{\Gamma^o}
\end{equation}
The activity coefficient is defined from the general equation (\ref{eq:mu_a}).  Identifying the two expressions, we obtain $\gamma_a=\Gamma_\mathrm{C}/\Gamma^o$. So as soon as the condensed phase appears, it deviates from the ideal value $\gamma_a=1$. At the end of the phase transition, the activity coefficient is then:
\begin{equation}
    \ln \gamma_a= \ln \left( \frac{\Gamma_\mathrm{C}}{\Gamma_\mathrm{V}}\right)
\end{equation}
With the values obtained here by the two fits, we obtain $\frac{\Gamma_\mathrm{V}}{\Gamma_\mathrm{C}}\approx 6$ and the corresponding free energy is $k_\mathrm{B}T \ln \gamma_a \approx -2 k_\mathrm{B}T$.  The problem is that it is difficult to estimate $\Gamma_\mathrm{C}$ because it corresponds to very dilute systems, which are not easily modeled by molecular simulations. It requires very large slab surface area and thus to simulate a very large number of water molecules. This is all the more important as the concentration range where the ideal law is found in the isotherm is not easy to estimate, given the error bars. In addition, from the first simulations with two bombykols on the surface, significant interactions are visible on the snapshots. It is therefore possible that, in reality, the result may be different if phase separation occurs earlier and the activity term reaches the threshold value of $14\ k_\mathrm{B}T$.

Phase separation may also occur earlier than in this modelling, for instance if crystallization of chains occur in the surface layer. The experimental investigation by X-ray reflectivity using a high-brilliance source at ESRF of a  macroscopic droplet  of alcohol in thermodynamic equilibrium  with a monolayer have suggested that a compact surface layer of alcohol has  a melting temperature  layer in equilibrium with saturated vapor pressure of up to 40~$^\circ$C higher than in the bulk \cite{Berge}. This has been experimentally verified even in the case of decanol, with a melting temperature of the bulk of 6~$^\circ$C and well above 40~$^\circ$C at the surface of clean water. Moreover, this work on shorter fatty alcohol than decanol has evidenced a discontinuous transition for longer chains, as in the modelisation presented here. It is therefore very likely that the Gibbs energy of absorption of compact 2D liquid condensed state in equilibrium with liquid bulk (equivalent to saturated vapor) has a free energy of binding of the order of 30~$k_{\mathrm{B}}T$. We had shown previously that adsorption of an isolated molecule on a clean surface of the order 10 kBT per molecule, not enough alone to  induce irreversible absorption on aerosol droplets \cite{Jami2022}. On the contrary, if crystallization occurs the gain in free energy could therefore be of 30 to 40~$k_{\mathrm{B}}T$ and adsorption of pheromone on aerosol surface in the form of condensed patches, with residence time, once adsorbed, of hours. This could explain the long residence times of low-volatility compounds on clean water surfaces, viewed not as a collection of isolated molecules, but rather as films \cite{Li,Woden}. Therefore, the aerosol transport that cannot occur for individual molecules on a clean surface is occurring by packets over long distances is possible via collective effects that are linked to phase transition in the form of condensed states of the surface layers of pheromones having a  low volatility and a vanishing solubility in water such as bombykol.

\clearpage
\section{Conclusions \& Perspectives}

Motivated by the important role played by physical chemistry and phase changes in pheromonal communication of insects, we specifically studied here the collective behavior of representative molecules adsorbed at air/water interface. Such interface serves as model interface for aerosol transport \cite{Jami2020} and we previously studied the behavior of single pheromones \cite{Jami2022}. The interest of this work is therefore manifold.

In terms of physical chemistry, the interfacial behavior of Bombykol might be a model for the behavior of many similarly structured molecules and our work is firstly a generic attempt to represent a lateral equation of state by molecular simulation. The proposed method, based on the calculation of surface tension, thus provides an understanding of the different adsorption regimes. At low concentrations, the molecules are parallel to the surface. However, when the molecules are more concentrated, they condense and turn upright until only the polar heads are in contact with the water. At the same time the monolayer stops following a linear Henry behavior, but instead likely undergoes a two-dimensional liquid-gas phase transition  between a condensing surface concentration of about $\Gamma_\mathrm{C} = 0.24\:$nm$^{-2}$ and a vaporisation concentration of $\Gamma_\mathrm{V} = 1.2\:$nm$^{-2}$. In our example, this seems to be fairly well described by the soft-sticky particle EOS model up to a surface concentration of $\Gamma_a = 1.4\:$nm$^{-2}$. However, the repulsive force is very weak and is modified for higher surface concentrations, as amphiphilic molecules can deform and reorient themselves. These crucial findings enable us to have a global and fundamental description of the behavior of non-model molecules like pheromones at the interface. In this sense, our work has also extended relevance for the field of atmospheric chemistry, where environmental gas-particules exchanges of low volatility molecules play a crucial role.

Secondly, when expressed in terms of chemical potential, these results highlight the very important role of the bombykol phase transition at the interface. It is in fact this transition, and in particular the condensing surface concentration $\Gamma_\mathrm{C}$, i.e. the surface concentration when the fluid phase appears, that controls the activity coefficient. In the context of bombykol interacting with aerosols, the proposed EOS fits yield to a $\Gamma_\mathrm{C}$ value not low enough to predict significant adsorption of bombykol on pure aqueous aerosols. However, this conclusion should be treated with caution for several reasons. On one hand,  the difficulties in modeling dilute surfaces using molecular simulations are real. This implies that  adsorption of this pheromone on pure aqueous aerosols might well occur, if the interactions between bombykol molecules condensed on the surface are found more favorable. On the other hand, aerosols in Nature are usually chemically heterogeneous and might contain many molecules, including fatty acids, which could ease the adsorption of Bombykol on the aerosol surface. 

This work can thus be pursued by specifically addressing this phase transition on more complex and hence realistic interfaces. Moreover grand Canonical Monte-Carlo simulations  could be used to  more properly address the initiation of the fluid phase. Biaised simulations can also be used \cite{park_calculating_2004, Jami2022} which could circumvent the high numerical cost of direct simulations, as simulating dilute systems would then not be necessary. Indeed, biaised simulations can estimate the free energy of adsorption (eq. \ref{eq:ads_coll}) of molecules directly on dense monolayers. However, the numerical cost of such a simulation method can instead increase if the relaxation kinetics of dense monolayer system is slow \cite{park_calculating_2004}. 

In terms of biological implications, two warrant further modeling and experimentation. Considering first the pheromone release process, pursuing the work with higher surface concentration would enable us to relate the results to the vapor pressure and surface tension of pure solutions of pheromones. High surface concentration system could also lead to the observation of multi-layering of the surfactants and Brunauer-Emmett-Teller (BET) adsorption equilibrium.  Vapor pressure and surface tension are important physico-chemical properties which have only recently been experimentally estimated for bombykol and a host of other moth pheromones, \cite{tiryaevaEstimatingSurfaceTension2024}, so time is ripe for further theoretical and computational developments.

In the context of pheromone detection by searching insects, the occurrence of a phase transition points to a strong threshold effect, whereby molecules non-linearly switch from dense to dispersed states when transported in the air at the surface of aerosols. The desorption of pheromones from the aerosol and subsequent adsorption onto the antennal cuticle—also covered by a patchy mosaic of liquid and semi-solid lipid phases \cite{kanekoCuticularLipidTopology2021, huthmacherImportanceBeingHeterogeneous2025}—thus also constitutes  interfacial phase change processes. To date, effects of pheromonal phase change have not been explicitly investigated, despite their central role in olfactory perception. Exploring such phenomena therefore opens a promising and original research avenue, with the potential to significantly advance our understanding of chemical communication.

\section*{Acknowledgment}
This work benefited from the high performance computing facilities of the CaSciModOT federation and the CERES3 facilities of the CEA center of Marcoule. This work was supported by the project France 2030, via the CNRS, projet PheroInnov (ANR-24-RRII-0001), and an ENS Lyon Ph.D. fellowship to L.J. in Tours, under the
supervision of J.C. and J.-F.D.

\clearpage
\bibliography{biblio.bib}

\end{document}